\documentclass[12pt,preprint]{emulateapj}
\usepackage{graphicx}

\usepackage{reaction}

\shorttitle{Monolithic Michelson for Precision Radial Velocity}
\shortauthors{Mahadevan et al.}

\begin{document}


\title{An Inexpensive Field-Widened Monolithic Michelson Interferometer for Precision Radial Velocity Measurements}


\author{Suvrath Mahadevan\altaffilmark{1}, Jian Ge, Scott W. Fleming, Xiaoke Wan, Curtis DeWitt, Julian C. van Eyken, \& Dan McDavitt}
\affil{Astronomy Department,  University of Florida, 211 Bryant
Space Science Center P.O. Box 112055 Gainesville, FL 32611-2055}
\email{suvrath@astro.ufl.edu}


\altaffiltext{1}{Visiting Astronomer, Kitt Peak National
Observatory, National Optical Astronomy Observatory. KPNO is
operated by AURA, Inc.\ under contract to the National Science
Foundation.}


\begin{abstract}
We have constructed a thermally compensated field-widened monolithic Michelson
interferometer that can be used with a medium-resolution
spectrograph to measure precise Doppler radial velocities of
stars. Our prototype monolithic fixed-delay interferometer is
constructed with off-the-shelf components and assembled using a
hydrolysis bonding technique. We installed and tested this
interferometer in the Exoplanet Tracker (ET) instrument at the
Kitt Peak 2.1m telescope, an instrument built to demonstrate the
principles of dispersed fixed delay interferometry. An iodine cell
allows the interferometer drift to be accurately calibrated,
relaxing the stability requirements on the interferometer itself.
When using our monolithic interferometer, the ET instrument has no
moving parts (except the iodine cell), greatly simplifying its
operation. We demonstrate differential radial velocity precision
of a few m s$^{-1}$ on well known radial velocity standards and
planet bearing stars when using this interferometer. Such
monolithic interferometers will make it possible to build
relatively inexpensive instruments that are easy to operate and
capable of precision radial velocity measurements. A larger
multi-object version of the Exoplanet Tracker will be used to
conduct a large scale survey for planetary systems as
part of the Sloan Digital Sky Survey III (SDSS III). Variants of the techniques and principles
discussed in this paper can be directly applied to build large
monolithic interferometers for such applications, enabling the
construction of instruments capable of efficiently observing many
stars simultaneously at high velocity-precision.
\end{abstract}




\section{INTRODUCTION}
Since the discoveries of the first extrasolar planets
\citep{Wolszczan92, Mayor95} more than 290 planets have been
discovered using radial velocity techniques, transit searches, and
microlensing. The majority of the known extrasolar planets have
been discovered using precision radial velocity measurements that
detect the gravitational influence of the planet on the parent
star. The presence of the planet induces a periodic change in the
line of sight velocity of the star and this velocity is measured
using the small Doppler shift in the spectral lines of the star.
Ongoing radial velocity surveys with high resolution echelle
spectrographs have achieved velocity precisions of $1-3$ m
s$^{-1}$ \citep{Butler96, Rupprecht04}, allowing the detection of $m\sin{i}=
7-20$ Earth mass planets close to their host star \citep{Santos04,
Rivera05}. These echelle spectrographs are generally large and
expensive instruments which are not easily duplicated, leading to
the availability of such instruments becoming a limiting factor in
intensifying high-precision radial velocity surveys. High precision radial
velocity observations of more stars are necessary in
order to discover lower mass planets, explore different regimes of
stellar mass and evolution (Johnson et al. 2007), planet formation
around young stars (Setiawan et al. 2007), and to discover more
planets around low-mass M stars (e.g. Endl et al. 2007).

Over the past few years we have developed the dispersed fixed-delay
interferometer (DFDI) technique \citep*{Erskine00, Ge02a, Ge02b}
into viable instruments capable of achieving high precision radial velocity
measurements on stars and detecting extrasolar planets. Such
instruments are substantially less expensive than echelle
spectrographs since they do not require high spectral resolutions
or large-format gratings. At the heart of these instruments is a
stable, field-widened, fixed-delay Michelson interferometer. One
such instrument, the Exoplanet Tracker (ET) instrument at the Kitt
Peak 2.1m has been used to confirm known exoplanets (van Eyken et
al. 2004) and detect a hot Jupiter around HD102195 (Ge et al.
2006). The ability to reliably construct monolithic Michelson
interferometers has the potential to make such instruments stable,
easy to use, and relatively easy to duplicate. This will enable
multiple versions of these instruments to easily be built to
complement existing echelle spectrographs in order to meet the
need for high precision radial velocity measurements to support upcoming space
missions like {\it KEPLER} (Borucki et al. 2003) and {\it SIM}
(Shao et al. 2007). An additional advantage of the DFDI technique
is the ability to observe a large number of stars simultaneously
using a wide field-of-view telescope. This technique will be
utilized to conduct MARVELS (Multi-object Apache point observatory Radial Velocity Exoplanet Large-area Survey), a multi-object survey for exoplanets,
as part of SDSS III (Ge et al. 2007). MARVELS will be able to
simultaneously monitor 120 F,G, and K stars in the 3 degree
field of view of the Sloan 2.5m telescope, efficiently searching
for Jupiter-mass planets \citep{Ge07}.

The Michelson interferometer is a crucial component of the ET and
MARVELS instruments and needs to be stable in order to reliably
recover relative velocity drifts of a few m s$^{-1}$. Phase drifts
of the interferometer during an exposure will tend to wash out the
fringes, reducing the measured fringe visibility and the velocity
precision. Large drifts between observing runs make it difficult
to achieve coherent velocity links over large timespans. The
current Michelson interferometer setup for the ET instrument at
Kitt Peak uses a piezoelectric transducer (PZT) to control the path difference. This stabilizes the
interferometer fringe and reduces the interferometer fringe phase
drifts. The need to run a closed-loop system on the most critical
component also makes observations difficult, since the observer
has to continually monitor the phase, and restart the PZT after a software or computer failure. In terms of a long duration survey for exoplanets, failure
of the PZT can lead to the inability to coherently link radial
velocities. The PZT control also becomes increasingly more difficult
for multi-object instruments based on ET that use much larger
beamsplitters due to the need to accommodate more fibers.

 To address some of these problems and to make ET-like instruments more
stable and substantially easier to use, we have been pursuing a
program to design and build an inexpensive monolithic
interferometer that satisfies our criteria of field widening and
stability (see Mahadevan et al. 2004). Monolithic interferometers
have also been discussed by Mosser et al. (2007) for use in
asteroseismology from DOME C in Antarctica. Polarizing Michelson
interferometers similar to our prototype have been in use for
solar oscillation measurements for a number of years in the Global Oscillation Network Group
(GONG, Harvey et al. 1995) and Solar and Heliospheric Observatory Michelson Doppler Imager (SOHO MDI)
instruments and are the design chosen for the Helioseismic and Magnetic Imager (HMI) instrument
(Graham et al. 2003) on the planned Solar Dynamics Observatory.
Monolithic interferometer assemblies have also been used to measure the solar resonance fluorescence of OH (Englert
et al. 2007). Polarizing interferometers are usually designed to
observe a single absorption line, while interferometers built for
spatial heterodyne spectroscopy are only effective over a small
wavelength region. The monolithic interferometer for the ET
instrument needs to be able to function in the wavelength range
5000-6000 \AA, be non-polarizing, and have a slight tilt in one
mirror to created the fringes observed with ET.

In this paper we describe the construction of an inexpensive
prototype monolithic interferometer and present results
demonstrating the ability to acquire precise stellar radial
velocities using the monolithic interferometer with the Exoplanet
Tracker (ET) instrument.

\section{Fixed Delay Interferometers: History and
Background}
  By the term fixed-delay interferometer we refer to interferometers
that operate in a narrow range of delays (usually a few $\mu$m
change in path length) unlike the interferometers used in Fourier
transform spectrographs which need to be scanned over large path
lengths. Fixed-delay interferometers and interferometers coupled
to spectrographs have often been used to aid calibration and
enable precise wavelength and velocity measurements. Fabry-Perot interferometers have been
used with spectrographs to measure precise radial velocities of
stars (McMillan et al. 1994; Connes et al. 1996) by using them to
track and calibrate out instrument drifts. Fixed-delay
interferometers are also routinely used for high precision
velocity measurements of solar oscillations (Gorskii \& Lebedev
1977; Beckers \& Brown 1978; Kozhevatov 1983; Harvey et al 1995
and Graham et al. 2003). Their use has also been proposed to
detect stellar oscillations and oscillations of Jupiter (Forrest
et al. 1978; Mosser et al. 1998; Mosser et al. 2003).

 For an
interferometer with an optical path difference $d$:
\begin{equation}
m \lambda=d
\label{eqn:delayfringe}
\end{equation}
where $m$ is the fringe order and $\lambda$ the monochromatic
wavelength of interest. A small change in the wavelength due to a
Doppler shift will also change the fringe order since the
interferometer delay ($d$) remains the same if the interferometer
is stable.
\begin{equation}
\Delta m=\frac{d}{\lambda^2}\Delta \lambda
\label{eqn:delayfringe1}
\end{equation}
Using the Doppler shift formula for non-relativistic velocities
\begin{equation}
\frac{\Delta \lambda}{\lambda} = \frac{\Delta v}{c}
\end{equation}
we can write Equation \ref{eqn:delayfringe1} as
\begin{equation} \Delta m = \frac{d}{\lambda} \frac{\Delta
v}{c} \end{equation} \begin{equation} \Delta v= \frac{c \lambda}{2
\pi  d} \Delta \phi \label{eqn:delayfringe2}
\end{equation}
 where $\phi=2 \pi m$ is the phase of the interference order.
  A change in velocity causes a Doppler shift in the wavelength of
the light. This Doppler shift can be measured from the phase shift
($\Delta \phi$) if the interferometer delay ($d$) is known. The
phase and phase shift can be measured either by stepping the
interferometer through small changes in the optical delay, or by
putting a small tilt on one interferometer mirror so that the
phase can be measured in a single exposure using a two dimensional
image.

\begin{figure}[htbp] \begin{center}
\includegraphics*[width=3in,angle=0]{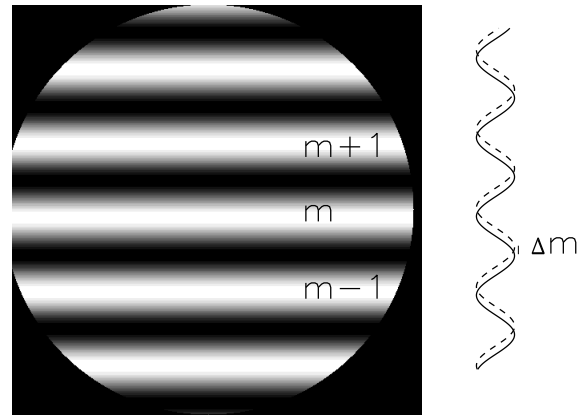}
\end{center} \caption[Phase shift of the interferometer fringes]
{This pattern shows the expected output when a field compensated Michelson interferometer with a small tilt in one arm is illuminated with monochromatic light. The cartoon on the right shows the phase shift of the interferometer fringes when the wavelength of
the light experiences a Doppler shift.}
\label{fig:interferometerfringe}
\end{figure}

Figure \ref{fig:interferometerfringe} illustrates the phase
pattern for a monochromatic emission line. The phase pattern shown
is for a collimated beam passing through a field-widened Michelson
interferometer that has a slight tilt on one mirror. A change in
the wavelength of the incident light leads to a shift in phase of the
output fringe pattern from the interferometer. The velocity shift
can then be estimated using Equation \ref{eqn:delayfringe2} since
the shift in phase of the sinusoidal pattern can be easily
measured. This ability to measure phase shifts is also applicable
for a narrow wavelength band containing a single absorption line
(Harvey et al. 1995). Real stellar spectra, however, have
thousands of stellar lines and using the largest possible number
of lines will yield the best velocity precision. A large
wavelength range of light passing through the interferometer will
lead to a dramatic reduction of the recorded fringe visibility
since different wavelengths will have different phase shifts in
the output pattern. Almost no interference fringes will be visible
in the output phase pattern, making the estimation of phase shifts
and velocity very difficult. Using an interferometer alone one can
achieve good velocity precision on bright sources, but the use of
only a single absorption line, or a small wavelength region, makes it difficult to obtain the velocity accuracy required to detect planets around stars due to the limited number of photons collected.

\begin{figure}[htbp] \begin{center}
\includegraphics*[width=3in,angle=0]{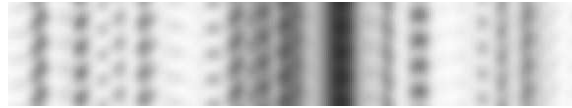}
\end{center} \caption[Two dimensional spectral format]{Simulated two dimensional spectral format for a dispersed fixed-delay interferometer instrument. Fringes are clearly visible at the location of the stellar absorption lines.} \label{fig:ETsimulspec} \end{figure}

      In 1997 David Erskine (then at Lawrence Livermore National Labs)
proposed using a field-widened Michelson interferometer in series
with a medium resolution spectrograph for the measurement of
precise radial velocities. The spectrograph is used to disperse
the light, allowing the measurement of fringe visibilities over a
significantly wider wavelength range. The first results from such
an instrument are found in Erskine \& Ge (2000) and Ge et al.
(2002), and the theoretical basis for the use of such an
instrument is outlined in Ge (2002), Erskine et al. (2003) and
Mosser et al. (2003). Figure \ref{fig:ETsimulspec} shows a
simulation of the spectral format expected from a dispersed fixed-delay
interferometer instrument. The interferometer fringes are clearly
visible due to dispersion by the medium-resolution
spectrograph. The normalized fringe pattern ($I$) in the slit
direction for each wavelength channel can be expressed in terms of velocity using
Equation \ref{eqn:delayfringe2}
\begin{equation}
I=1 + \gamma \sin {(2\pi d \frac{\Delta v}{c\lambda} +\phi_0})
\label{eqn:a}
\end{equation}
where $\phi_0$ is the phase of the first pixel and $\gamma$ is the
fringe visibility defined using the maximum ($I_{max}$) and
minimum ($I_{min}$) intensities of the sinusoidal fringe pattern
\begin{equation}
\gamma = \frac{I_{max} - I_{min}}{I_{max} + I_{min}} \label{eqn:b}
\end{equation}

A Doppler velocity shift of the absorption line manifests itself
as a phase shift of this sinusoidal pattern. The accuracy with
which the radial velocity can be measured from one fringe is
related to the photon noise error and the fringe visibility ($\gamma$) and
is discussed in \S 4.2.

\section{The Current Exoplanet Tracker Instrument}

\begin{figure}[htbp]
\begin{center}
\includegraphics*[width=6in,angle=0]{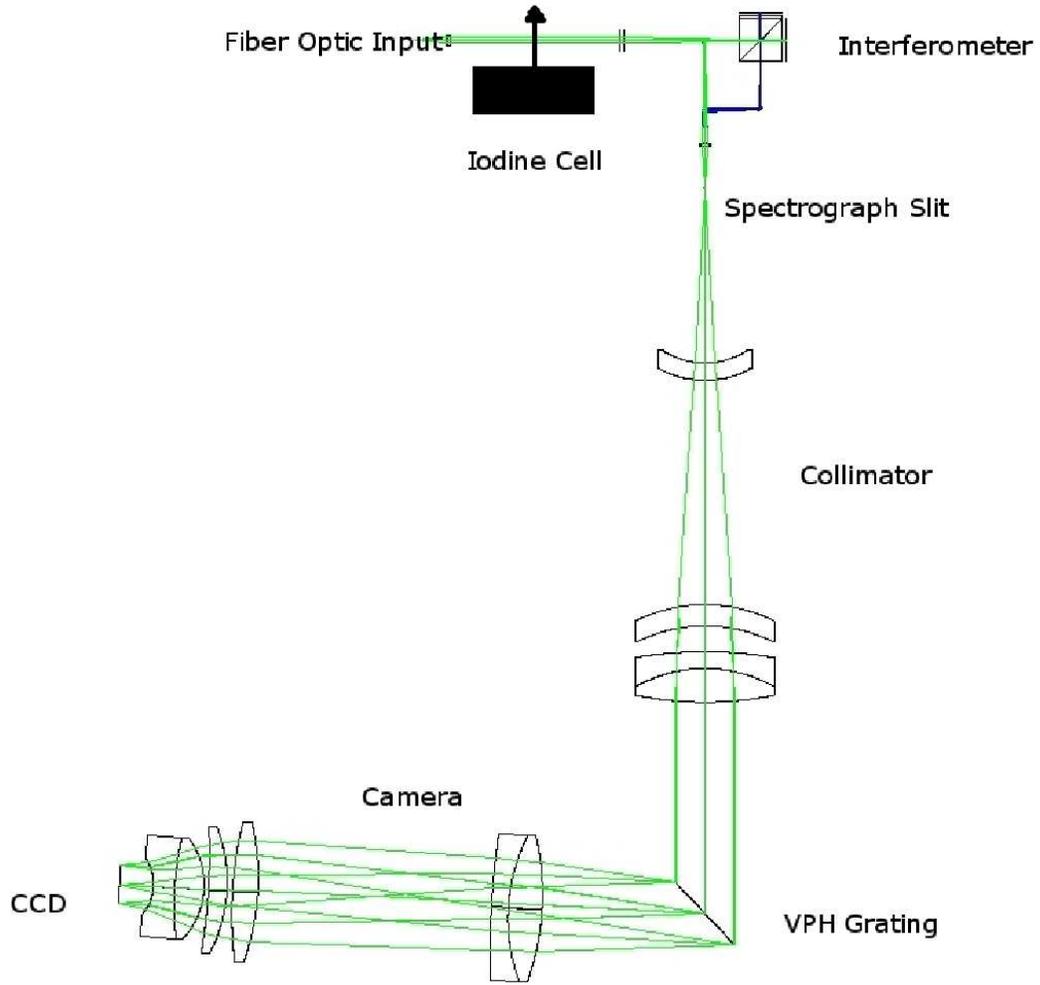}
\end{center}
\caption[Optical layout of the Kitt Peak ET instrument]{Optical
layout of the Kitt Peak ET instrument, showing the Michelson
interferometer, spectrograph camera and collimator, VPH grating,
and the iodine cell that can be inserted into the path of the
collimated beam.} \label{fig:EToptical}
\end{figure}

The current Exoplanet Tracker (ET) instrument at Kitt Peak is the
end product of many years of experimentation with various design
ideas for high throughput DFDI instruments.
ET has been built to enable precision radial velocity measurements on
slowly rotating main sequence stars, and the exoplanet HD102195b
was discovered using radial velocities obtained with this
instrument (Ge et al. 2006). Aspects of the instrument design have
been mentioned in van Eyken et al (2003, 2004),
Ge et al. (2003, 2004, 2006), and Mahadevan et al. (2004, 2008). In this section we
briefly review the design aspects of the instrument that are
relevant from the point of view of installing and testing a
prototype monolithic interferometer. Figure \ref{fig:EToptical}
shows the optical layout of the KPNO ET instrument. Relevant
components are described below.

\subsection{Optical Fibers and Fiber Feed}
  The optical fibers and fiber feed used in ET are inherited from the Fiber Optic Echelle instrument (Ramsey \& Huenemoerder
1986). The efficiency and properties of these optical fibers are
described in Barden (1998). Each fiber cable is enclosed in a
flexible metal covering and is 22 meters in length. The optical
fiber used in the ET instrument is a $200$ $\mu$m high $OH$ fiber.
In the f/8 beam of the telescope this fiber subtends an angle of
$\sim 2.5$ arc seconds on the sky. The use of the optical fiber
enables the ET instrument to be mechanically de-linked from the
telescope and in an isolated enclosure with very few moving parts.
This allows the instrument environment to be well controlled and
eliminates mechanical flexure due to movement, minimizing velocity
drifts that need to be calibrated out. The ET instrument can be
used with either the 2.1m telescope or the 0.9m Coud\'e
telescope simply by switching the 200$\mu$m optical fiber from
one telescope to another. The scrambling properties of the optical
fiber (Heacox 1986, Hunter \ Ramsey 1991) make the illumination
of the Michelson interferometer and spectrograph slit much more
uniform.

\subsection{Instrument Location and Enclosure}
\begin{figure}[htbp]
\begin{center}
\includegraphics*[width=4in,angle=0]{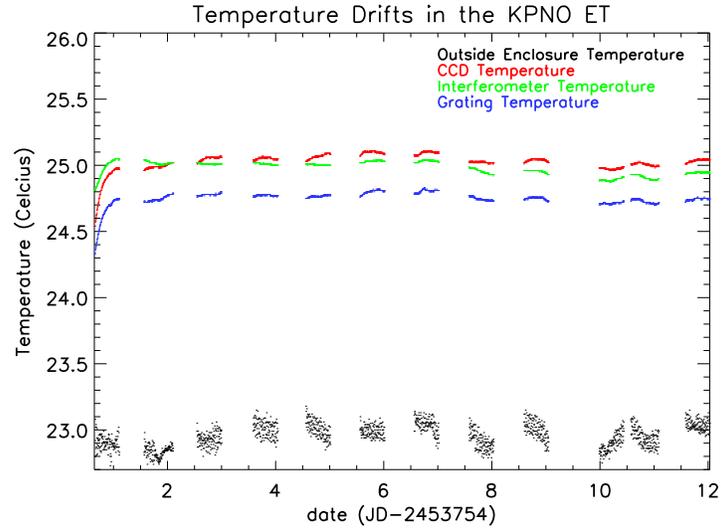}
\end{center}
\caption[The temperature measured at four points in the ET
instrument]{The temperature measured at four points in the ET
instrument during a typical observing run. The instrument
enclosure helps stabilize the temperature and dampens the effect
of the temperature variations outside the enclosure. The iodine
cell and the electronics inside the detector head are the dominant
source of heat inside the enclosure.} \label{fig:kpnotemperature}
\end{figure}

  The ET instrument is currently installed in a small insulated room within the
coud\'e spectrograph room at the base of the 2.1m telescope dome.
The spectrograph room is very stable and its temperature only
responds slowly to the outside temperature. The instrument
itself is built on a vibration isolated optical table. A baseboard heater is installed at one end of the room
and a thermostat at the other end senses the ambient temperature
of the air and controls the heater. A secondary enclosure made
of insulating panels is built on the optical table to further
reduce the effect of temperature variations and to minimize the
turbulence. Figure \ref{fig:kpnotemperature} shows the temperature
of the ET instrument measured at four locations. The temperature
of the instrument within the inner instrument enclosure is more
stable than the temperature of the room itself.

\subsection{Iodine Cell}
   An iodine cell manufactured by Triad Technologies is used in the ET
instrument to provide a simultaneous calibration of the instrument
drift. A motor allows the iodine cell to be inserted or removed
from the collimated stellar beam path. The cell is operated at a
temperature of $60 \pm 0.1^\circ$C. The dense absorption
spectra of $I_2$ are superimposed on the stellar absorption lines.
Since the starlight passes through the iodine cell, the two
intermingled spectra experience exactly the same illumination in
the interferometer and spectrograph, and therefore track the same
instrument drifts.

\subsection{Input Optics and Michelson Interferometer}
  The ET input optics are designed to account for the focal ratio degradation of the 200$\mu$m optical
fiber and able to collect most of the photons exiting the  fiber. The output beam from the fiber is collimated to a
diameter of $\sim 6.6$ mm and passes through the iodine cell if
the cell is inserted into the beam path. A cylindrical lens then
collapses the beam in one direction (dispersion, or horizontal
direction) and the beam is focused onto the mirrors of the
Michelson interferometer. The Michelson interferometer used in ET
has a 4 mm BK7 glass etalon in one arm, an air-gap in the other
and is designed to have an optical delay of $\sim 7$ mm. The
lengths of the two arms are also chosen so that the interferometer
optical delay is relatively insensitive to the angle of the
converging beam from the cylindrical lens. This `field widening'
of the interferometer is described in more detail in \S 4.3. A
small tilt in one beam of the interferometer allows the creation
of $3-4$ horizontal fringes over the 6.6 mm stellar beam. A slight
tilt of the interferometer beam cube allows both the transmitted
and reflected beams to be imaged onto the spectrograph slit using
relay mirrors.

\subsection{Piezoelectric Transducer for Interferometer Stabilization}
 We use a piezoelectric transducer (PZT) to keep the interferometer from
drifting during an exposure (drifts lead to decreased fringe visbility) and keep the delay stable. Piezoelectric materials can
generate mechanical stress in response to applied electric
voltage, enabling the interferometer delay to be maintained even
in the presence of vibration and temperature drifts. The PZT is
attached to one of the interferometer mirrors and is controlled
using a LabView program. A single-mode fiber fed with a stabilized
He-Ne laser (632.8 nm) is used to illuminate a small portion of
the beam splitter. The horizontal fringes formed are recorded by a
video camera and are used as the input to the LabView program. The
position of these fringes is then locked by applying the
appropriate voltages to the PZT. This method can effectively
stabilize the interferometer fringe pattern over the period of an
observing run ($7-12$ days). To obtain flats and calibration lamp
spectra the PZT voltage is ramped very quickly, leading to the PZT
mirror vibrating and washing out the interference fringes. The PZT
is turned off after every observing run, re-initialized and dialed
back to its previous position at the start of a new observing run.

\subsection{Spectrograph}
 The ET spectrograph is a transmissive design with an f/5 collimator, a
volume phase holographic transmission grating and a $\sim f/2$
camera. The spectrograph optics were designed by Dequing Ren. The
stellar beam from the interferometer is imaged on the spectrograph
slit and collimated. Unlike normal spectrographs, where the
starlight is brought to a focus at the slit, the ET slit is not at
the point where the beam is smallest but where its fringe
visibility is maximal. The pre-slit optics form an image of one of
the interferometer mirrors onto the slit, and change the slow beam
to an f/5 beam. The spectrograph then disperses the light, forming
an image of the slit (and the interference fringes) on the
detector.

\subsection{Spectral Format \& Efficiency}
\begin{figure}[htbp]
\begin{center}
\includegraphics*[width=5in, height=5in, angle=90]{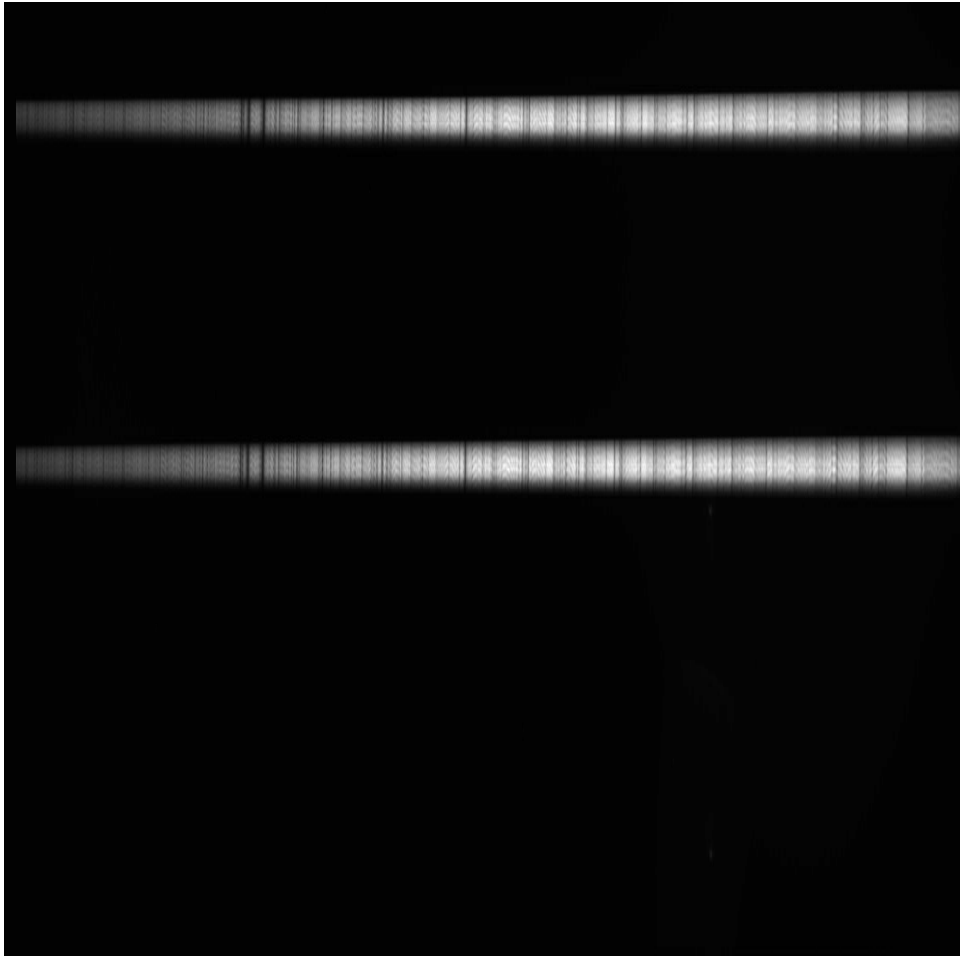}
\end{center}
\caption[The spectra format of the ET instrument]{The 4k $\times$ 4k pixles image obtained with ET instrument. Both outputs from the interferometer
are imaged onto the detector. The two-dimensional spectral format
allows the measurement of the fringe phase at each wavelength
without having to physically change the interferometer delay.}
\label{fig:ET_format}
\end{figure}

 The ET instrument operates in the grating first order, but its spectral
format is two dimensional to allow the measurement of the fringe
phase at each wavelength. The $250$ $\mu$m slit is sampled by 6.7
pixels on the detector, yielding a spectral resolution of $\sim
6000$. The fringes in the slit direction are sampled by $\sim 58$
pixels. Figure \ref{fig:ET_format} shows the spectral format of
the instrument. The top and middle spectra are the two outputs
from the interferometer. The measured instrument efficiency from
the fiber output to the detector plane is $\sim 49 \%$. This
efficiency includes losses from the lenses, mirrors,
interferometer, slit and the grating, but not the iodine cell or
the CCD. The CCD is expected to have a quantum efficiency of $\sim
90\%$ in our wavelength range. Using the measured fiber and
telescope transmission in average seeing, and assuming their
efficiency to be 35-40$\%$ in average seeing the total instrument
efficiency is calculated to be $\sim$ 15.5 -- 17.5 \%. The
instrument efficiency in good seeing condition, measured using a
flux calibration star, is $\sim 18\%$ and in good agreement with
the expected values.

\section{A Monolithic Michelson Interferometer for ET}
 A monolithic interferometer for an instrument like ET has a
 number of requirements in terms of delay, field widening and
 stability. In this section we explore these parameters and the
 materials that can be used in the construction of a monolithic
 interferometer.
\subsection{The Choice of Optical Delay}
The spectral format of ET allows one to measure the phase shift of the
interference pattern and convert that to velocity. As outlined in
Butler et al. (1996) and \citet{Ge02a}, the velocity
uncertainty from one pixel of the normalized fringe is:

\begin{equation}
\sigma_{RMS}= <\frac{\epsilon_{I}}{dI/dv}>
\end{equation}

The uncertainty in the measured intensity, $\epsilon_{I}$, is just
the inverse of the signal to noise ratio (S/N$^{-1}$) if the S/N
is large enough to ignore the dark current and read noise from the
detector, a condition that always needs to be satisfied to achieve
high-precision radial velocities with ET. For the sinusoidal interferometer
fringes observed with ET:
\begin{equation}
<\frac{dI}{dv}>= \frac{\sqrt{2}\pi d \gamma}{c \lambda}
\end{equation}
The radial velocity error from measurements of one fringe is then
given by
\begin{equation}
\sigma_{RMS}=\frac{c\lambda}{\sqrt{2}\pi d \gamma
\sqrt{N_{photons}}}
\end{equation}

where $N_{photons}$ is the total number of photons recorded for
all the pixels in the interferometer fringe. The functional form
of this equation means that the highest velocity precision for a
given number of photons is obtained when the value of $d \gamma$
is maximized. The value of the interferometer delay that yields
the largest $d \times \gamma$ for a given wavelength region is, however,
a function of the spectral type and the rotational velocity of the
star. Since the interferometer effectively heterodynes information
to lower spatial frequencies (Erskine et al. 2003), a fast rotator
has less high frequency information and requires a smaller delay.

\begin{figure}[htbp] \begin{center}
\includegraphics*[width=5in,angle=0]{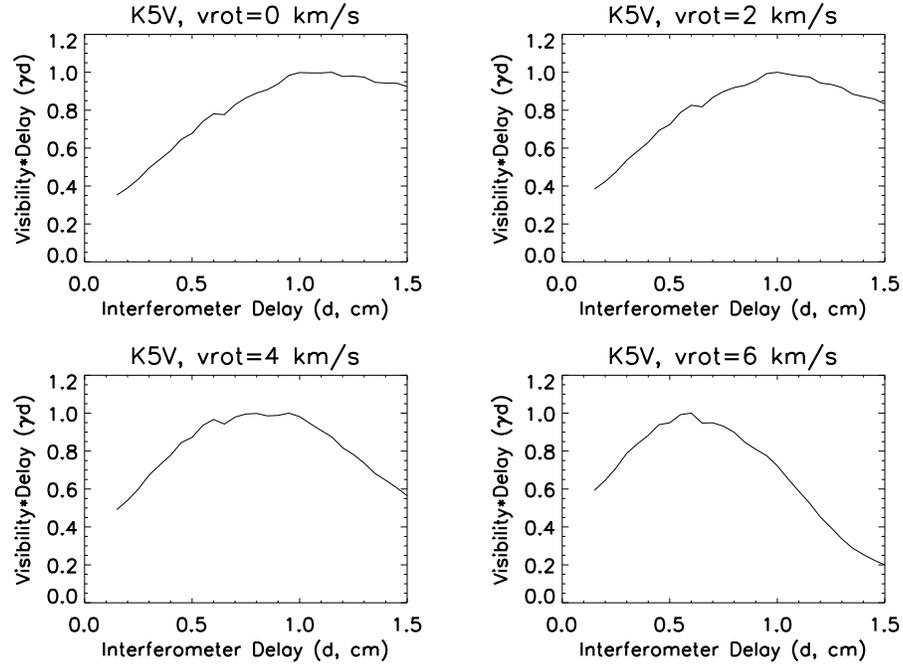}
\end{center} \caption[Optimal delay for a K5V star]{Optimal delay for a K5V star at different rotational velocities. The optimal value of  the interferometer delay maximizes $\gamma d$. This maximum value of $\gamma d$ has been normalized to one for all the plots.}
\label{fig:rotators} \end{figure}

Figure \ref{fig:rotators} shows the $d \gamma$ at different
optical delays ($d$) for a K5V star at various rotational velocities.
These results are derived from simulations of the ET instrument
spectral format using stellar models as an input. The chosen delay
for the ET instrument is $\sim 7$ mm since this value provides
adequate fringe visibility ($\gamma$) for late F, G and K stars
that have $0< v \sin{i} < 12$ km s$^{-1}$. Fast rotators (e.g. the
planet bearing star Tau Boo) have very low fringe visibility at
our chosen interferometer delay, making precise radial velocities
difficult to extract. The optical delay of the interferometer does
not need to be exact for our application, It merely needs to be
stable. More detailed discussions regarding
the calculation of optimal delay for different spectral parameters
can be found in Mosser et al. (2003).

\subsection{The Field Compensation Principle}

  The optical delay of a Michelson interferometer is strongly
dependant on the incident angle of the light. Deliberately
designing an interferometer to decrease this dependence of the
optical delay on the incident angle is known as field widening, or
field compensation, and is described in more detail in Hillard \&
Sheperd (1966), Shepherd et al. (1985), and Thullier \& Sheperd
(1985). A Michelson interferometer for ET must have arms of
different lengths to create a total optical delay of $\sim 7$ mm,
i.e., $\sim 14000$ waves of delay at a wavelength of 5000 \AA.
This delay is appropriate for observations of stars with $v
\sin{i} < 12$ km s$^{-1}$. For an interferometer with arm lengths
$L_1$ and $L_2$, with refractive indices $n_1$ and $n_2$
respectively, the interferometer optical delay for an on-axis ray
($d_0$) is given by

\begin{equation}
 d_0=2(n_{1} L_{1} - n_{2}L_{2})
\label{eqn:delay}
 \end{equation}

For a ray entering the Michelson interferometer at an angle $i$
the interferometer delay can be written as (Shepherd et al. 1985)

\begin{equation}
d = 2[n_{1}L_{1}(1-\frac{\sin^{2}i}{n_{1}^{2}})^{-\frac{1}{2}} -
n_{2}L_{2}(1-\frac{\sin^{2}i}{n_{2}^{2}})^{-\frac{1}{2}} ]
\end{equation}

Expanding all terms of $\sin{i}$ upto fourth order and ignoring
higher order terms we get
\begin{equation}
d = 2[(n_{1}L_{1} - n_{2}L_{2}) - \frac{\sin^{2} i}{2}
(\frac{L_{1}}{n_{1}} - \frac{L_{2}}{n_{2}}) - \frac{\sin^4 i}{8}
(\frac{L_1}{{n_1}^3} -\frac{L_2}{{n_2}^3})] \label{eqn:c}
\end{equation}

If the same material is used in both arms of the interferometer
($n_1=n_2$) the interferometer delay can be quite sensitive to the
angle of the incident light. Such an interferometer is often
referred to as an uncompensated interferometer. For example, an
interferometer with air gaps in both arms and a 7 mm delay (at
5000 \AA) has 19.1 waves of phase change over a 3 degree
half-angle and 4.5 waves over a
1.5 degree half angle. Such large phase changes are unacceptable with the ET
instrument since the fringe visibilities will be substantially
reduced when the fringes are imaged on the slit.

The $\sin^2{i}$ term in Equation \ref{eqn:c} for the delay
disappears when

\begin{equation}
\frac{L_1}{n_1}=\frac{L_2}{n_2} \label{eqn:compensation}
\end{equation}

The delay for a compensated Michelson interferometer (ie. one that satisfies Equation \ref{eqn:compensation}) is then given by:

 \begin{equation}
d= d_0[1 +\frac{\sin^4 i}{8{n_1}^2{n_2}^2}]
\end{equation}

where $d_0$ is the interferometer delay when $i= 0$, ie. when the
ray is along the axis of the interferometer. For small values of
$i$ the delay is almost independent of the incident angle. Further
increase in field widening (so called super field widening) can be
achieved by using two or more glasses with an air space to also
reduce the effect of the $\sin^4{i}$ term (eg. Gault, Johnston \&
Kendall 1985), but we have not attempted this for our prototype monolithic interferometer.

Any field compensated Michelson interferometer at a specific delay must therefore simultaneously satisfy Equations
\ref{eqn:delay} and \ref{eqn:compensation}. The complete
disappearance of the $\sin^2{i}$ term in Equation
\ref{eqn:c} can only be achieved for a specific design
wavelength (5000 \AA\ in this case). The simplest choice for the ET interferometer was to use the same material as the beamspliiter (BK7) in one arm and an air-gap  in the other. To find the values of $L_1$
and $L_2$ that satisfy our constraints we calculate the refractive
index of BK7 at the design wavelength using the Sellmeier dispersion formula. The
dispersion constants used in this equation may vary slightly
depending on the glass melt and the values we use are for typical
melts of N-BK7. The values of the arm lengths that satisfy the
field widening condition are $L_1$(BK7) = 4.04 mm and $L_2$(air) =
2.66 mm. The wavelength dependence of the refractive index and
errors in the dimensions of the arms will, in reality, lead to a
situation where the compensation condition is not met exactly. To
evaluate the performance in such a case we define a term
$\epsilon$ which is a wavelength dependant measure of the
departure from the field widening condition.

\begin{equation}
\epsilon=\frac{L_1}{n_1}-\frac{L_2}{n_2}
\end{equation}

\begin{figure}[htbp]
\begin{center}
\includegraphics*[width=4in,angle=0]{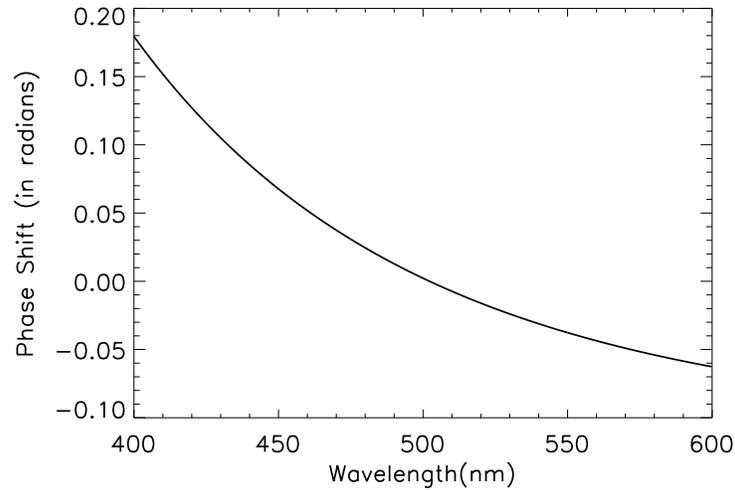}
\end{center}
\caption[Phase shift for a compensated Michelson
interferometer]{Phase shift for a compensated Michelson
interferometer that has $L_1$(BK7) $=4.04$ mm, $L_2$(air) $=2.66$
mm. The wavelength dependent phase shift show here is for a
converging beam with a half angle of 1.5 degrees.}
\label{fig:phaseshift}
\end{figure}

Even when $\epsilon=0$ at 5000 \AA\ the variation of refractive
index with wavelength results in a non-zero value of $\epsilon$
creating a small dependence of the delay on the $\sin^2{i}$ term
for wavelengths other than 5000 \AA. The phase difference caused by
this effect at the calculated values of $L_1$ and $L_2$ is
illustrated in Figure \ref{fig:phaseshift} for a beam with a
half-angle of 1.5 degrees. Although the ET instrument only works
in the 5000-5600 \AA\ region, such an interferometer can, in
principle, be used over wider wavelength regions, and we have
calculated the phase difference for 4000-6000 \AA\ which is the
spectral range containing most of the radial velocity information
for F, G, K stars \citep*{Bouchy01}. For a hypothetical case in
which the compensation condition is off by $\epsilon=100$ $\mu m$
even at 5000 \AA\ , a phase shift of 0.137 waves occurs for a beam
with a half angle of 1.5 degrees. This will lead to a slight
decrease in the observed visibility, but is otherwise acceptable
as long as the focal ratio or the illumination pattern of the beam
does not change significantly. If an iodine cell is used in the
beam path, then small changes in the illumination do not matter
since the iodine and starlight are both affected in the same way.
The scrambling properties of the optical fibers used in ET also
ensure that the illumination of the interferometer is stable even
in bad seeing conditions.

\subsection{Reducing Sensitivity to Temperature Drifts}
     The need for a particular value of the optical delay and field
compensation fixes the values of the arm lengths $L_1$ and $L_2$
since we have decided to use BK7 glass in one arm and an air-space
in the other. However, since our goal is to build a stand alone
monolithic interferometer, whose arms are isolated from any
external mechanical mounts, the choice of spacer material for the
air arm still needs to be made. The delay for a beam going through
a compensated Michelson interferometer with a very small angle of
inclination is shown in Equation \ref{eqn:delay}. Following
\citet*{Title80}, variation of the interferometer phase with
temperature can then be written as

\begin{equation}
\frac{d (d_0)}{d T}=2(n_1 \frac{d L_1}{dT} + L_1\frac{dn_1}{dT}-n_2 \frac{d L_2}{dT} -L_2\frac{d n_2}{dT})
\end{equation}

Using the field widening condition (Equation
\ref{eqn:compensation}) the right hand side of this equation can be recast in the form

\begin{equation}
2\frac{L_1}{n_1}({n_1}^2(\frac{1}{n_1}\frac{dn_1}{dT}+\frac{1}{L_1}\frac{dL_1}{dT})-{n_2}^2(\frac{1}{n_2}\frac{dn_2}{dT}+\frac{1}{L_2}\frac{dL_2}{dT}))
\label{eqn:athermal}
\end{equation}

which disappears when

\begin{equation}
{n_1}^2(\frac{1}{n_1}\frac{dn_1}{dT}+\frac{1}{L_1}\frac{dL_1}{dT})={n_2}^2(\frac{1}{n_2}\frac{dn_2}{dT}+\frac{1}{L_2}\frac{dL_2}{dT})
\end{equation}

    This thermal compensation condition depends only on the refractive
indices, their variation with temperature, and the linear thermal
expansion coefficients of the materials used, and is independent
of the actual arm lengths or path differences. In our case the
left-hand-side of the equation is already predetermined by the
choice of BK7 as the material in that arm. A wise choice of spacer
material for the other arm can then be used to athermalize the
interferometer. To evaluate the variation of refractive index with
temperature of BK7 we use the equation

\begin{equation}
\frac{dn_{abs}}{dT}= \frac{n^2(\lambda, T_0)-1}{2n(\lambda, T_0)}
(D_0+2D_1\Delta T+3D_2{\Delta T}^2 +\frac{E_0 +2E_1\Delta
T}{\lambda^2-\lambda_{TK^2}})
\end{equation}


where $\lambda$ is the wavelength of interest, $\Delta T$ is the
temperature difference from the reference temperature $T_0$ ($20^\circ$ C) and $n$ is the refractive index for that wavelength
calculated at the reference temperature using the Sellmeier
dispersion formula. Using this equation, the variation of
refractive index of BK7 with temperature can be calculated with
constants ($D_0-D_2, E_0, E_1$) provided in the Schott optical glass catalog. Using
Equation \ref{eqn:athermal} and assuming vacuum operation, the
change in delay and the velocity drift can be estimated for
various materials. Using Equation \ref{eqn:delayfringe2}, the
shift of one fringe ($2\pi$ in phase) corresponds to a velocity
shift of
\begin{equation}
\Delta v = \frac{c \lambda}{d_0}
\end{equation}

For our value of $d_0 \sim 7$ mm, one fringe shift corresponds to a
velocity shift of $\sim 21.5$ km s$^{-1}$. Table
\ref{tbl:velshift} shows the velocity shift for a one degree
change in temperature using BK7, copper, CaF$_2$, or aluminium
as a spacer material for an interferometer operating in vacuum.
CaF$_2$ proves to be the best spacer but is difficult to work with
since it is fragile. For assembling our prototype interferometer
we chose copper as a spacer material since it provides a
significant improvement from a BK7 spacer, and is easy to work
with. We have chosen to design the interferometer for vacuum
operation since this significantly increases the stability by
removing the effect of the refractive index of air changing due to
pressure and humidity variations (J. Sudol, private
communication).\footnote{http://gong.nso.edu/instrument/instrument\_performance/vres/vres.html}
While designed for operation in vacuum, this interferometer can
also be operated in air, although the effect of a one degree
temperature increase will rise from $\sim 250-500$ m s$^{-1}$ to
$\sim1000$ m s$^{-1}$. This still provides a significant
improvement from using a BK7 spacer in air or vacuum. The ET
instrument is temperature stabilized and the temperature of the room at the location
of the interferometer rarely changes beyond $0.3^\circ$ C and is often stable to less than $0.2^\circ$ C over
the course of an observing run (see Figure
\ref{fig:kpnotemperature}). An interferometer built with a copper
spacer would therefore be expected to be stable at a level of
$\sim$200 m s$^{-1}$ in air over an observing run if the observed
velocity drift is purely due to the interferometer delay changing
as a function of temperature.

\section{Assembling a Monolithic Fixed-Delay Interferometer}

\subsection{Components}
\begin{figure}[htbp]
\begin{center}
\includegraphics*[width=3in,angle=0]{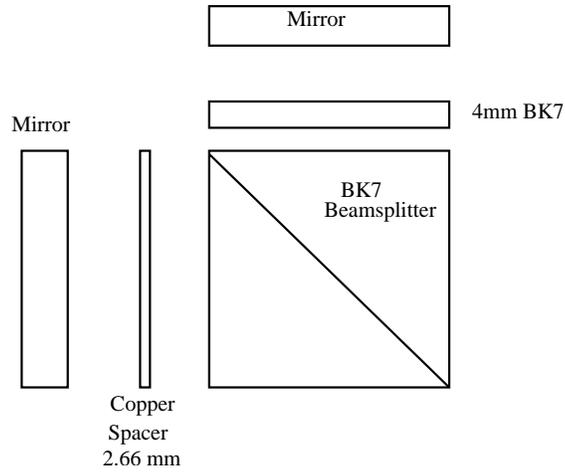}
\end{center}
\caption[Phase shift for a compensated Michelson
interferometer]{Exploded view of the monolithic Michelson
interferometer showing each component of the final assembly.}
\label{fig:exploded}
\end{figure}

 We used commercially available optical components
to assemble the monolithic interferometer prototype since the goal
was to determine if a relatively inexpensive interferometer can be
used in a DFDI instrument to
derive precise radial velocities. The components used in the
assembly were a Newport 1 inch broad-band non-polarizing beam
splitter, an uncoated BK7 window of diameter
1 inch and thickness 4 mm from Edmund Optics to
use as the optical delay in one arm, and 1 inch mirrors from
Thorlabs. The Newport beam splitter cube had a proprietary
broad-band anti-reflection coating on it made of alternate layers
of TiO and SiO. 
A copper ring of thickness 2.66 mm was also machined for use as a
spacer material in the air arm. Small holes were drilled around
the surface of the copper ring to enable the evacuation of air so
that the final assembled interferometer could also be operated or
tested in a vacuum vessel. Figure \ref{fig:exploded} shows each component in an exploded view of the final
interferometer assembly.

\subsection{Bonding Technique}
 To assemble these components into a single interferometer we used
 a hydroxide catalyzed bonding procedure developed by Jason Gwo at Stanford University to meet the high stability requirements of the
Gravity Probe B mission (Gwo 1998a,b). Such optical bonding
techniques are also currently being explored for use in the LISA
space mission. This bonding technique can link two surfaces by
forming a silicate-like network through hydroxide catalyzed
hydration and dehydration. The bond that is formed is transparent
to visible wavelengths, and expected to be very stable. The
hydroxide catalyzed bonding can work at room temperature, and
interface thickness are expected to be less than $200 \mu$m. The
technique can be applied to any  surface capable of forming a
silicate-like network. For such a surface (eg. fused silica, BK7,
alumina) the equations for the hydration and dehydration in the
presence of a hydroxide catalyst is \citep{Gwoa, Gwob}
\begin{equation}
Si-O-Si + H_20 \reactionrevarrow{KOH}{} Si-OH + Si-OH
\end{equation}

\begin{figure}[htbp]
\begin{center}
\includegraphics*[width=4.6in,angle=0]{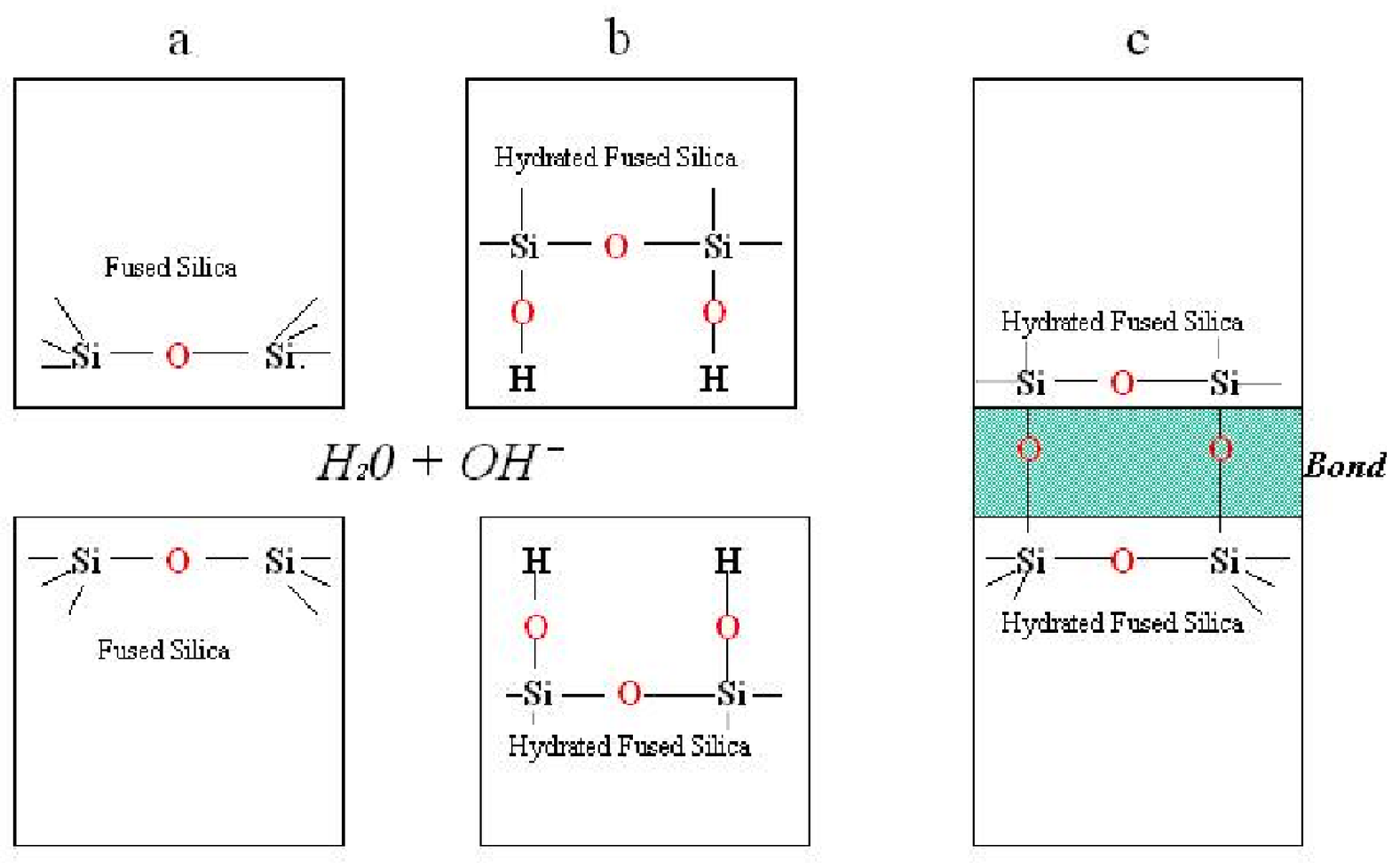}
\end{center}
\caption[The formation of a bond between two fused silica
disks]{The formation of a bond between two fused silica disks in
the presence of water and the hydroxide catalyst.  A similar
mechanism occurs for any material capable of forming a
silicate-like network. In this figure a) shows the fused silica
disks, b) demonstrates the formation of hydrated fused silica, and
c) shows the formations of a bond linking the two disks. This
figure is adapted from \citet{Gwob}.} \label{fig:optobond}
\end{figure}

The bonding process is illustrated for fused silica in Figure
\ref{fig:optobond}. The presence of the hydroxide ion catalyzes
the formation of hydrated fused silica, which is then able to form
a bond, linking the two disks together. As the two disks are
allowed  to cure at room temperature the loss of water from the
interface leads to a stronger bond. The two components can be
manipulated during the bond cure time to achieve the desired
optical alignment. This bonding technique is applicable to
materials that can both form silicate-like networks, and also if
one of the materials can generate a silicate-like network and the
other is capable of forming a layer of hydroxyl groups (-OH). This
allows a number of metals (aluminium, brass, copper etc.) and
metal oxides to be bonded to glasses capable of forming
silicate-like networks (eg. BK7, fused silica).

\subsection{Assembly}
The optical components were cleaned by immersing them in acetone
and agitating in an ultrasonic bath for five minutes. To eliminate
any residue after the acetone cleaning the components were also
cleaned by immersing them in iso-propanol and de-ionized water
(DI) and agitated again in the ultrasonic bath. The copper
ring was immersed in acetone for 24 hours to remove grease from
the machining process and then cleaned using the same procedure as
that followed for the optical components. The entire assembly of
the interferometer was performed in a clean room. To assemble the
BK7 arm of the interferometer we mixed a solution of KOH of 1:500
molar ratio. To prevent contamination by particulate matter the
KOH solution was applied to the optical surfaces of the mirror
using a syringe with a 0.2 $\mu$m particulate filter attached to
its tip. The 4 mm BK7 glass window was then placed on the mirror,
and more KOH applied to the exposed surface. Finally the
beamsplitter was placed on the assembly. One of the advantages of
this bonding procedure is that the bonding is not instantaneous,
and small alignments are possible. The three pieces were then
aligned together, with the weight of the beamsplitter keeping the
three components in contact till they were bonded. An optically
transparent bond was achieved in $48-72$ hours and the pieces could
not be separated by hand.

Bonding the copper ring to the mirror to create the air arm of the
interferometer was more challenging. One side of the
copper ring was polished with various grades of polishing paper
until interference fringes of the required density and tilt were
observed when the copper ring and mirror were placed on the
beamsplitter assembly and observed with a 543 nm He-Ne laser. Copper cannot form a silicate-like
network, and the machined copper ring was not optically flat. To
bond the copper to the mirror we used a sodium silicate solution.
The silicate in this solution allows for easier bonding and can
fill in small gaps between the pieces. Optical transparency of the
bond was not an issue here, since the copper ring is only a spacer
for the air arm. A small amount of sodium silicate solution was
placed on the copper ring using the tip of a syringe and the
mirror was gently lowered onto the ring. Unlike KOH, the sodium
silicate solution left substantial residue, and we were careful
not to let it get on the exposed mirror surface. The copper spacer, now
bonded to the mirror, was then tested again with the beamsplitter
to ensure that the fringes were present. A small amount of further
polishing and alignment of the copper spacer was required to ensure that the correct
fringe pattern was still observed.

\begin{figure}[htbp]
\begin{center}
\includegraphics*[width=3.6in,angle=90]{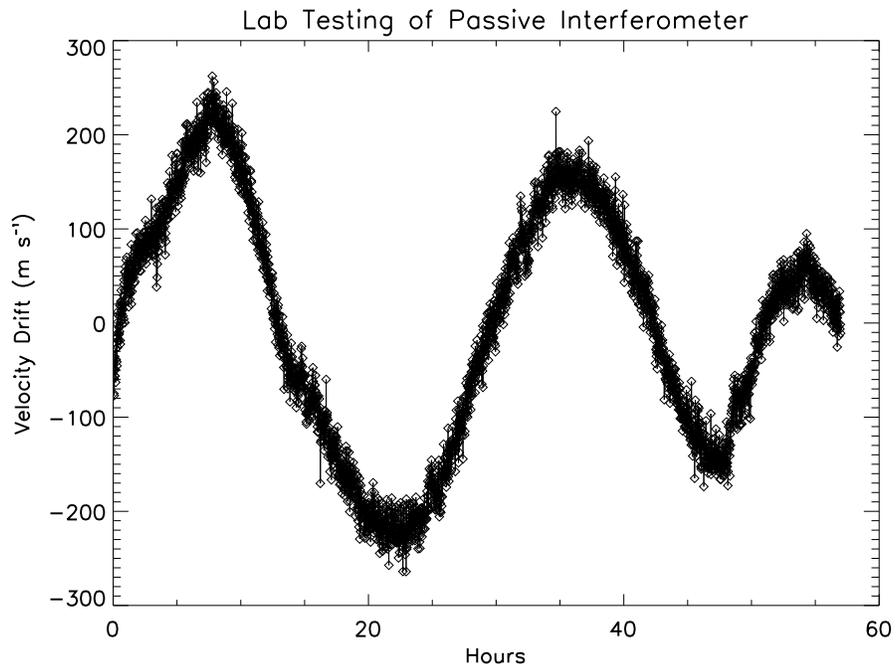}
\end{center}
\caption[The velocity drift of a passive interferometer during lab
testing]{The velocity drift of a passive interferometer during lab
testing. The interferometer was placed in a vacuum chamber and the
fringes were monitored using a stabilized He-Ne laser and a
LabView program.} \label{fig:labtest}
\end{figure}

We were apprehensive about
using sodium silicate solution to bond the copper ring to the
beamsplitter, and so we used UV curing optical glue (NOA 73). The
glue was applied to the copper ring, and a UV lamp was used to
flash cure the glue when the correct optical alignment was
achieved. The entire assembly was then allowed to settle for 48
hours before further testing.
  Three prototype interferometers were assembled using the procedures
described above. One of the interferometers was installed in the
Kitt Peak ET instrument, and its performance is described in more
detail in the following section. Another interferometer was placed
in a custom designed vacuum enclosure on a vibration isolated
table, and its fringe pattern was monitored for more than 50 hours
using a stabilized He-Ne laser (632.8 nm). Since the optical delay
of the interferometer is known, the phase shift of the
interferometer pattern can be converted to a measured velocity
shift using Equation \ref{eqn:delayfringe2}. The results of this
test are shown in Figure \ref{fig:labtest} and demonstrate that
our prototype monolithic interferometer is capable of achieving a
velocity stability of $\sim 500$ m s$^{-1}$ over a few days. The
interferometer vacuum enclosure itself was not temperature
controlled and we suspect that the quasi-periodic variations in
velocity are caused by diurnal variations in temperature. Such
variations are quite acceptable, since the fringe shifts due to instrument drifts will be very small over a typical stellar exposure using ET (15-20 minutes), and will not impact the observed visibilities.

\section{On-Sky Results with the Monolithic Interferometer}
\begin{figure}[htbp]
\begin{center}
\includegraphics*[width=3.6in,angle=0]{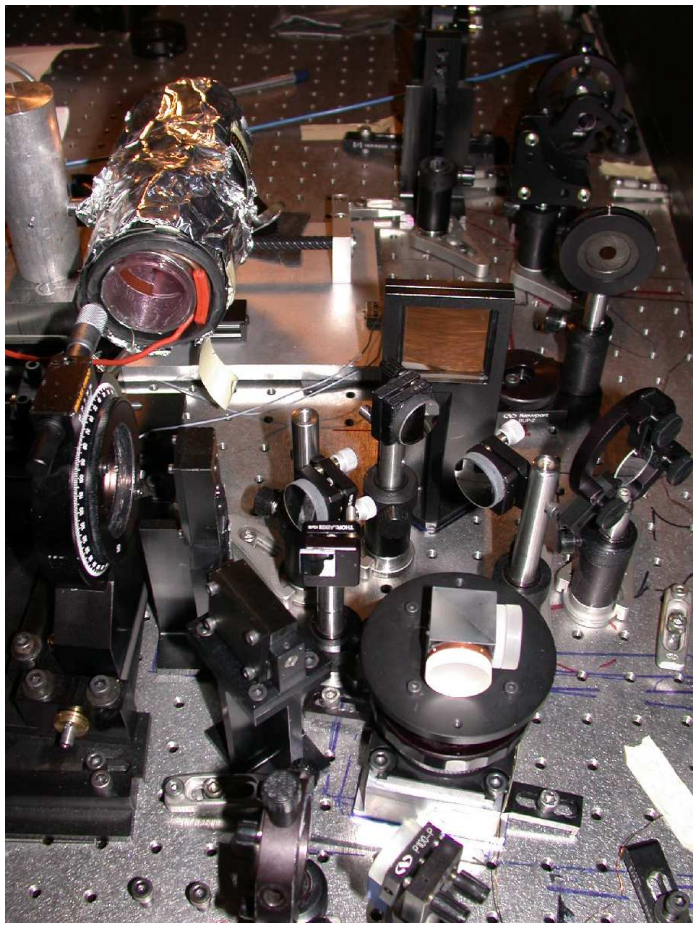}
\end{center}
\caption[The assembled monolithic interferometer aligned in the
Kitt Peak ET instrument]{The assembled monolithic interferometer
aligned in the Kitt Peak ET instrument. The copper spacer and the
BK7 arm of the interferometer are clearly visible. Also seen are
the iodine cell, the input optics and the slit assembly.}
\label{fig:passivesetup}
\end{figure}

One of the monolithic interferometers was transported to Kitt Peak
National Observatory and installed in the ET instrument in May
2005. The monolithic interferometer replaced the existing actively-controlled interferometer (with its PZT mirrors and mounts) for
the duration of this observing run. Figure \ref{fig:passivesetup}
shows the passive interferometer set-up with the ET instrument.
The input optics, iodine cell and the slit assembly are also
clearly visible in the figure.

\subsection{Observations}
 Once the monolithic interferometer was installed and aligned with ET we were
able to observe a number of stars using the 0.9m Coud\'e
telescope. During the observing run (29th May 2005 through 6th
June 2005) we obtained multiple data points on stars known to be
stable over the short term, such as $\eta$ Cas (V=3.45 mag) and 36 UMa
(V=4.83) as well as stars with known planetary companions (55 Cnc,
$\rho$ Crb). The standard observing procedure with this instrument
is described in Ge et al. (2006) and Mahadevan et al. (2008), and
we mention it only briefly here. Individual star and iodine
templates were obtained for each star, and each star was then
observed with the iodine cell inserted in the path of the stellar
beam. The iodine spectrum acts as a fiducial for determining and
subtracting off the instrument drift. A number of additional spectra of the
iodine cell (illuminated by a quartz lamp) were also observed over
the duration of the run. This iodine cell data allows one to
characterize the instrument drift during the run. To verify the
performance of the passive interferometer a single mode fiber, fed
with a stabilized He-Ne laser (632.8 nm), was used to illuminate a
small portion of the passive interferometer and this phase pattern
was continuously recorded by a video camera. This phase pattern
allowed us to continuously monitor the total drift of the passive
interferometer over the entire observing run.

\subsection{Data Analysis}
The data acquired with the monolithic interferometer could not be
processed automatically with the standard ET data pipeline because
the standard calibration sequence of non-fringing tungsten lamp flats and
non-fringing Thorium-Argon (Th-Ar) spectra were not acquired.
These are usually acquired with the ET instrument by
jittering the PZT very fast to wash out the interferometer
fringes. Since the monolithic interferometer was deliberately
designed to have no moving parts such flats cannot be obtained. The tungsten
lamp flats are used to correct pixel-to-pixel variations and
partially correct for illumination variations, and the
non-fringing Th-Ar emission lamp spectra are used to determine the
slant of the spectra in the slit direction. Not correcting the
pixel-to-pixel variations may lead to excess radial velocity
noise, and correcting the slant is essential to determine
accurate velocity shifts.

We were unable to reliably measure the
slant from fringing Th-Ar spectra and we decided to use the slant
correction parameters from the preceding observing run. The
observed slant is primarily caused by distortions introduced by
the spectrograph optics, and by the positioning of the slit. The
use of the slant solution from a previous run (May 27 2005) is
justified since we were careful to not move the spectrograph, and
to maintain the same slit optics and beam paths while replacing
the actively controlled interferometer with the monolithic
interferometer. Using the slant solutions from the previous run
enabled radial velocities to be extracted using the standard ET
data processing pipeline.

The pipeline processing steps are described in more detail in van
Eyken et al. (2004), Ge et al. (2006), and Mahadevan et al. (2008).
The data produced by ET were processed using standard IRAF
procedures, as well as software written in Research System Inc.'s
IDL software. The images were corrected for biases, dark current,
and scattered light and then trimmed, illumination corrected,
slant corrected and low-pass filtered. The visibilities ($V$, same
as $\gamma$) and the phases ($\theta$) of the fringes were
determined for each channel by fitting a sine wave to each column
of pixels in the slit direction (see Figures 2, 3). To determine
differential velocity shifts the star+iodine data can be
considered as a summation of the complex visibilities (${\bf
V}=Ve^{i\theta}$) of the relevant star ($V_Se^{i\theta_{S_0}}$)
and iodine ($V_Ie^{i\theta_{I_0}}$) templates (van Eyken et al.
2004; Erskine 2003, van Eyken et al. 2007). For small velocity
shifts the complex visibility of the data, for each wavelength channel, can be written as

\begin{equation}
V_D e^{i\theta_D} = V_S
e^{i\theta_{S_0}}e^{i\theta_S-i\theta_{S_0}}
+V_Ie^{i\theta_{I_0}}e^{i\theta_I-i\theta_{I_0}}
\end{equation}

where $V_D, V_S,$ and $V_I$ are the fringe visibilities for a
given wavelength in the star+iodine data, star template and iodine
template respectively, and $\theta_D, \theta_{S_0}$, and
$\theta_{I_0}$ the corresponding measured phases.
 In the presence of real velocity shifts of the star and instrument
 drift, the complex visibilities of the star and iodine template best match the
 data with phase shifts of $\theta_S - \theta_{S_0}$ and $\theta_I -
 \theta_{I_0}$ respectively. The iodine is a stable reference and the iodine phase shift tracks the instrument drift. The
difference between star and iodine shifts is the real phase shift
of the star, $\Delta \phi$, corrected for any instrumental drifts

\begin{equation}
\Delta \phi = (\theta_S - \theta_{S_0}) - (\theta_I -
\theta_{I_0})
\end{equation}

This phase shift can be converted to a velocity shift, $\Delta v$,
using Equation \ref{eqn:delayfringe2}. The ET data analysis
pipeline finds the shift in phase of the star and iodine templates
that are a best match for the data, and uses these phase shifts to
calculate the velocity shift of the star relative to the stellar
template.


\subsection{Interferometer Drift \& Environmental Stability}
Our experimental setup allowed us to measure and assess the
performance of the interferometer and its stability in two
different ways:

\begin{itemize}
\item{By monitoring the phase drift of a fringe pattern generated
using a using a stabilized He-Ne laser (632.8 nm). This fringe
pattern is generated by shining the laser light into the
interferometer via a single-mode optical fiber on a separate path. This fringe
pattern is continuously recorded by a video camera, and the phase
variation can be converted into velocity shifts.}
\item{By tracking the velocity drifts of the iodine templates
obtained throughout the observing run. The iodine in the
temperature-stabilized cell is very stable by design, and any measured
velocity shifts have to be a result of drifts in the
interferometer or spectrograph. }

\end{itemize}

\begin{figure}[htbp]
\begin{center}
\includegraphics*[width=3.6in,angle=90]{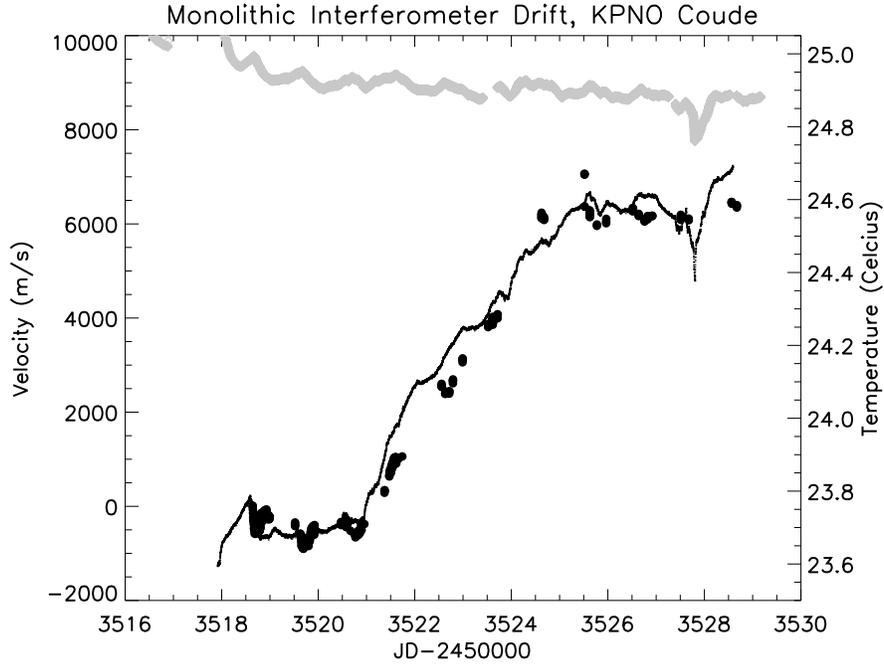}
\end{center}
\caption[The velocity drift of the monolithic interferometer
during an observing run at Kitt Peak]{The velocity drift of the
monolithic interferometer during an observing run at Kitt Peak, as
measured with a stabilized He-Ne laser, is shown as solid black line.
Filled circles are the measured drift of the iodine calibration
spectra obtained during the run. The grey line shows the
measured interferometer temperature.} \label{fig:kpnodrift}
\end{figure}

The results of these measurement techniques are expected to be
highly correlated, but not identical since they monitor different
parts of the instrument, use different wavelength regions and
different data analysis techniques are used. The interferometer
drift measured from the phase patterns recorded by the video
camera is shown in Figure \ref{fig:kpnodrift} as a solid line, and
the measured iodine velocity drift is plotted as filled circles.
These two velocities are forced to the same zero point at the
beginning of the observing run. The dotted line shows the
temperature measured at the location of the interferometer. The
passive interferometer drifted by $\sim 7000$ m s$^{-1}$ during
the observing run. The sharp dip in the temperature and the
velocities seen towards the end of the observing run was caused
when the oxygen alarm was triggered for the coud\'e spectrograph
room. Safety considerations required the door of the coud\'e
spectrograph room to be opened to the cold night air to ensure the
safety of the observer. Although the ET instrument room was not
opened, the cold air rushing into the telescope basement caused a
temperature dip in the ET instrument room. The prototype monolithic interferometer tested with the Kitt Peak ET exhibits
substantially more drift than the one tested in the vaccum
chamber. While some of this drift can be attributed to the
temperature and pressure shifts in the instrument enclosure,
Figure \ref{fig:kpnodrift} demonstrates that the temperature does
not shift enough to cause such a large drift. We suspect that a
significant part of this drift is caused by a variation in the
tilt due to slow warps of the copper ring, but the exact cause of
the drift is not trivial to determine. We surmise that
this drift was a consequence of internal stresses in the copper
ring being released as the interferometer attempted to achieve
equilibrium with the thermal environment.

\subsection{Radial Velocity Results}

\begin{figure}[htbp]
\begin{center}
\includegraphics*[width=3.6in,angle=90]{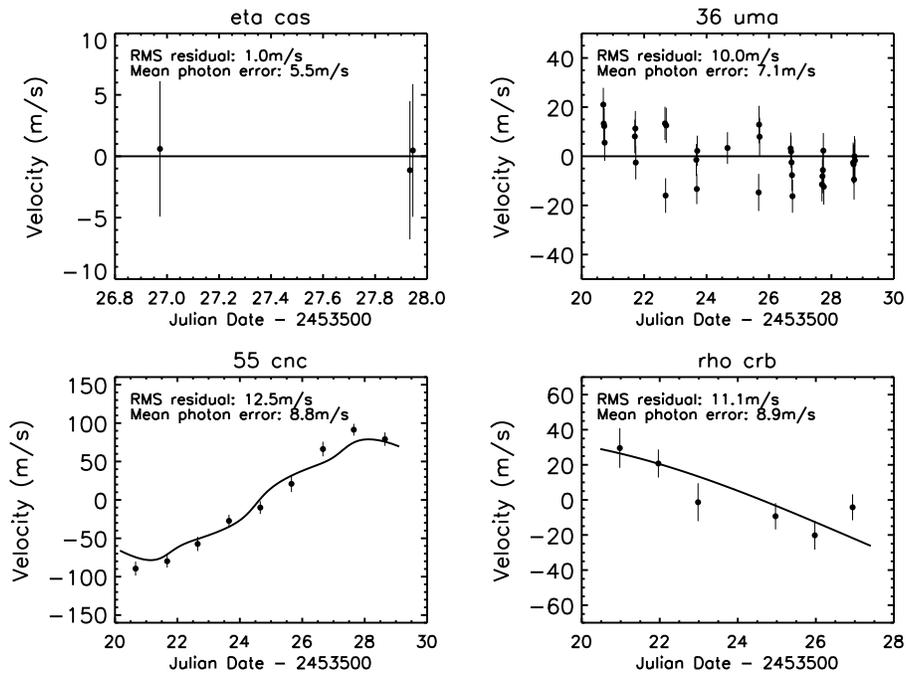}
\end{center}
\caption[Radial velocity results from the ET instrument and the
passive interferometer]{Radial velocity obtained with the
monolithic interferometer. The stars selected are the known radial
velocity stable stars $\eta$ Cas  and 36 UMa , and the known
planet bearing stars 55 Cnc and $\rho$ Crb. The expected velocity
amplitude caused by the known planetary systems is plotted as a
solid line.} \label{fig:velocities}
\end{figure}

  The radial velocities obtained for our velocity reference and planet
stars are shown in Figure \ref{fig:velocities}. The mean photon noise error for $\eta$ Cas is $\sim \sigma=5.5$ m s$^{-1}$,
and the data points exhibit a very low scatter only because of
chance alignment. $\eta$ Cas is a known long period binary whose
radial velocity is expected to be quite stable in the short term.
For 55 Cnc and $\rho$ Crb the expected radial velocity curves of
the stars due to the planets orbiting them (using orbital
parameters from \cite{Butler06}) are plotted as a solid line. The monolithic
interferometer drift of $\sim 7$ km s$^{-1}$ over the 12 day
observing run yields a mean long term drift of $\sim 24$ m
s$^{-1}$ hour$^{-1}$, which is acceptable since such a small drift
does not significantly affect fringe visibilities for even the
longest exposures which are typically ~20 minutes. The use of the
iodine cell for simultaneous calibration allows us to effectively
determine and subtract out the instrumental drift, yielding precise radial velocities.

\section{Discussion}
Using off-the-shelf components and a novel bonding technique we
have built a monolithic fixed-delay Michelson interferometer for
the ET instrument. With this inexpensive prototype we have
effectively demonstrated the ability of such an interferometers to
recover precise radial velocities. The results for the stable
and planet bearing stars demonstrate that the prototype
monolithic interferometer, developed using only off-the-shelf
components, is capable of allowing us to achieve a velocity
precision of $\sim 10$ m s$^{-1}$ or better over observing runs as
long as two weeks. Monolithic interferometers are significantly
more insensitive to vibrations, and do not need active locking
using a PZT device. The temperature-compensated, field widened
interferometer also makes the ET instrument significantly easier
to use since the observer no longer has to constantly monitor the
phase locking software.
  The performance of one of our prototype interferometers in the lab is
 substantially better than the one tested at KPNO. We believe that
 the large drifts seen at Kitt-Peak were caused by the warping on the copper ring. Nevertheless, we have demonstrated the ability of such a monolithic
 interferometer to aid in the precise measurement of radial
 velocities using a suitable reference to track the instrument and
 interferometer drifts.  The hydrolysis catalyzed bonding technique we have
 employed is transparent over a wide wavelength region, and stable
 over a large range in temperatures.
  The next goal in our development efforts is a larger and very stable
 field-widened and temperature-compensated fixed-delay
 interferometer that can be used in a multi-object instrument to
 observe many stars simultaneously. We expect that the issues related to the warping of the ring can be avoided if a second glass material were used instead of an airgap. Better designs of such
 interferometers, coupled with better environmental control, may
 make it possible to build interferometers that are intrinsically stable enough to not
 require simultaneous iodine calibration in order to measure to radial velocities to a high precision. With high enough stability and good thermal control it may become possible to achieve acceptable velocity precision by bracketing the stellar observations with a velocity reference, or using another reference fiber to track the star fiber. With minor variations, such interferometers can
 potentially be employed in a variety of dispersed fixed delay
 interferometer instruments including large multi-object
 instruments planned for upcoming surveys like MARVELS.

\acknowledgments  We thank Dequing Ren and Bo Zhao for useful
discussions and their contributions to the ET project. Jerry
Friedman shared his expertise in mechanical design for tests of an
earlier version of this interferometer. We are grateful to Richard
Green, Skip Andree, Daryl Wilmarth and the KPNO staff for their
generous support. This work is supported by National Science
Foundation grant AST 02-4309, JPL, the Pennsylvania State
University, and the University of Florida. JCvE., SM. and SWF. acknowledge travel support from KPNO. JCvE. and SM.
acknowledge the Michelson Fellowship. SWF was supported by the Florida Space Grant Fellowship. This research has made use
of the SIMBAD and Vizier databases, operated at ADC, Strasbourg,
France. This work was performed in part under contract with the
Jet Propulsion Laboratory (JPL) funded by NASA through the
Michelson Fellowship Program. JPL is managed for NASA by the
California Institute of Technology. This work is based on data
obtained at the Kitt Peak 2.1 m telescope.


\clearpage

\begin{table}[htbp]
  \begin{center}
    \caption[Velocity shift of the interferometer]{Velocity shift of the interferometer, for a one degree Celsius drift in temperature, when different materials are used as spacers. The numbers shown here are for an interferometer operating in vacuum. }
     \label{tbl:velshift}
    \begin{tabular}{lllllll}
      \hline
SPACER MATERIAL  &  VELOCITY SHIFT (m s$^{-1}$ degree$^{-1}$)\\
\hline
  BK7       &             $\sim$ 2700\\
  Copper      &          $\sim$ 250-500\\
  CaF$_2$     &           $\sim$ 50\\
  Aluminum    &          $\sim$ -1000\\
\hline
 \end{tabular}
  \end{center}
\end{table}

\clearpage

\end{document}